TUAP052

# GPIB ADDRESS CONVERTER
Y. Suzuki, M.Ieiri, Y.Katoh, E.Kusano, M.Minakawa, H.Noumi, M.Takasaki, K.H.Tanaka, and Y.Yamanoi
KEK, Tsukuba, Ibaraki 305-0801, Japan
## Abstract

A GPIB address converter (GAC) has been constructed. This paper reports on the function and test results. The GAC has two GPIB connectors (upper and lower ports). The upper port has a GPIB primary address, and is connected to a GPIB system controller. The lower port acts as a GPIB controller of the lower side GPIB line. The GPIB system controller can access the lower side GPIB devices through the GAC by using an extended two-byte address function. The two-byte address (primary + secondary) is shown in the combination of the GAC address and the address of the lower side device. The GAC converts the secondary address into the primary address of the lower side GPIB device. By using 30 GACs, the GPIB system controller can access 930 devices assigned only primary addresses.

## 1 INTRODUCTION

In controlling and monitoring the accelerator or the experimental physics, the GPIB is one of the useful field buses. When the primary address is used, one GPIB controller can control 30 devices. According to the specifications: (IEEE-488.1), one GPIB controller can control 960 devices when the extended two-byte address function is used. However, there is one inconvenience. There are many useful instruments equipped with GPIBs: oscilloscopes, multi-meters, and accelerator control devices. However, they are scarcely equipped with the extended two-byte address function. In this situation, the GPIB address converter (GAC) was developed. Figure 1 shows a photograph of the GAC. Figure 2 shows an example configuration of the GAC in a GPIB system.

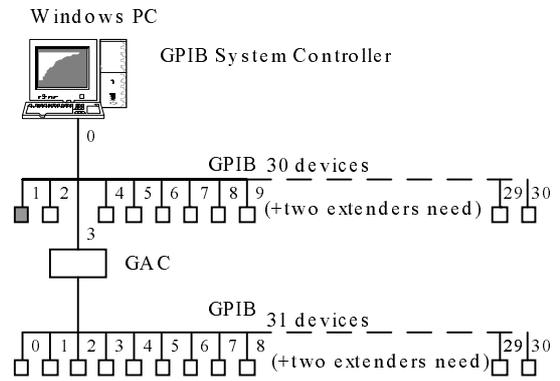

Figure 2. Configuration example

Figure-3 shows a block diagram of the GAC.

When the GPIB controller addresses the GAC, the GAC converts the secondary address into the primary address, and then the GAC controls the devices with the primary address through the lower GPIB port. The data, which pass through the GAC, do not change. Thus, the GPIB system controller does not need to add any program or to change. By using of this GAC, one GPIB controller can control 930 devices assigned only a primary address.

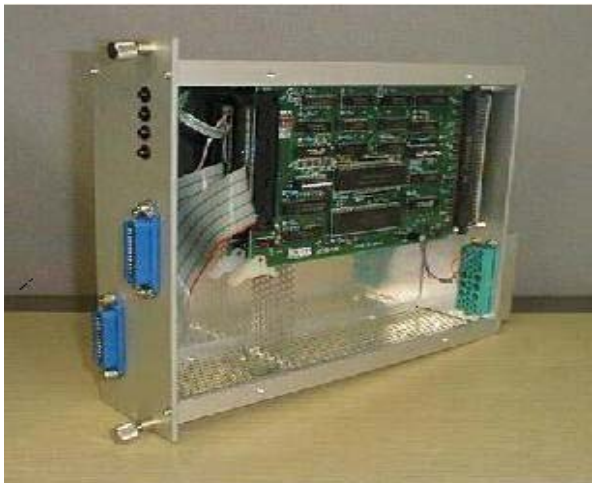

Figure 1. The GAC is assembled into a NIM module.

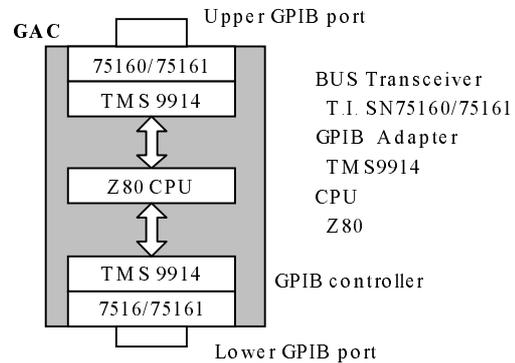

Figure 3. Block diagram of the GAC

## 2 GAC

### 2.1 Hardware

The main parts of the GAC are two GPIB adapters, TMS-9914 and an eight-bit microprocessor Z80. In the block diagram of Figure 3, the upper port acts as a GPIB device, not a controller, and occupies one GPIB primary address number between 0 and 30, except for the address number of the system controller. The lower port GPIB adapter acts as a GPIB controller of the TMS-9914's local mode, and thus it does not occupy any GPIB address. Thus, 31 numbers of 0 to 30 can be assigned to the lower side devices.

In an actual use situation, the maximum number of the GPIB devices directly connected to the GAC's lower port is limited to 14. The reason is based on the fan-out ability of the bus-driver: 75160/75161 (regulated by IEEE488). A GPIB extender is available for expanding of the bus.

### 2.2 Address Conversion

The secondary address number, received from the system controller, is converted into the primary address by following equations:

LAn: Listener primary Address n code
   codes 20H to 3EH are assigned to LAn 0 to 30
   codes 3Fh is assigned to UNL: Un-Listen
TAn: Talker primary Address n code
   codes 40H to 5EH are assigned to TAn 0 to 30
   code 5FH is assigned to UNT: Un-Talk
SCm: Secondary Address m code
   codes 60H to 7FH are assigned to SCm 0 to 31
1) In the listener condition
   LAn=SCm-40H
2) In the talker condition
   TAn=SCm-20H

Then n=m.

LAn and TAn are used to control of the lower side device.

### 2.3 Listener Sequences

The GAC receives the following message bytes from the system controller:
[Upper GPIB port]
**UNL** : inhibits all current listeners (Un-Listen)
**TAD** : system controller is assigned to the talker
**LAD** : GAC is assigned to the listener
**SCm** : receive secondary address
At this point, the GAC accepts all SCms.

The GAC sends the following messages to the lower side GPIB port:
[Lower GPIB port]
**UNL** : Un-Listen
**LAn** : LAn=SCm-40H

At this point GAC checks the upper GPIB port. Received interface messages are transferred to the lower side GPIB. Those messages are as follows:
**SDC** : Selected Device Clear
**GTL** : Go To Local
**GET** : Group Execute Trigger

When a **DATA** byte is received, the lower GPIB adaptor, TMS-9914, is set in the Talk Only mode, and the **ATN** (attention) signal is set to L (0). Then, the data read out from the upper GPIB port is repeatedly written in the lower GPIB port. **EOI** (end or identify) signal is also checked, and it is sent to the lower GPIB port with the data.

### 2.4 Talker and Serial Poll Sequences

The GAC receives the following message bytes from the system controller:
[Upper GPIB port]
**UNL** : Un-Listen
**LAD** : system controller is assigned to the listener
**TAD** : GAC is assigned to the talker
**SCm** : receive secondary address

The next step has two cases: one is **SPOLL** (Serial Poll); the other is a data request. However, it is not possible to know the serial pole in advance. Then, every time SPOLL is executed, STB is read out from the lower GPIB device. The flow is as follows:
[Lower GPIB port]
**UNL** Un-listen
**SPE** : serial poll enable
**TAm** : the lower device is assigned to Talker
**STB** (SBN or SBA): Status byte
**SPD** : serial poll disable
**UNT** : Un-talk

The STB from the device is set to the serial poll register of the upper GPIB adapter (TMS-9914).

Then, the operation of the upper GPIB is permitted (by the Z80-cpu), and the operation is continued. When SPOLL has not happened by the system controller, STA is stored in the STB memory for the next SPOLL (for when receive SCm). Then, the memory is cleared after SPOLL has been received.

(STA: represent a status byte sent by a device in which a request for service is indicated (bit 7=1)).

(STB: represents a status byte sent by a device in which a request for service is not indicated (bit 7=0)).

[Upper GPIB port]

**ATN=0** means a demand for DATA
The next procedures are as follows:
[Lower GPIB port]
**UNL** : Un-Listen
**TAm** : lower side devise is assigned to Talker
Set GPIB adapter TMS-9914 listen-only mode
**ATN=0**
**DATA** : read data are repeatedly sent to the upper GPIB port. The EOI signal of the lower GPIB port is checked, and is also sent along with the DATA byte.

*2.5 SRQ, SDC, DC, and IFC,*

The SRQ line in the lower GPIB port is always watched, and the SRQ signal is transferred immediately to the upper GPIB port. When STA is stored in the STB memory, an SRQ signal is also set.

The lower GPIB port executes SDC (selected device clear), DC, or IFC(Interface Clear) when those interface messages are received in the upper GPIB port.

## 3. OPERATION RESALT

To confirm the operation of the GAC, used equipment are shown in Table-1.

| Device | Address |
|---|---|
| GAC | 3 |
| Multi-meter HP34401A | 3+22 |
| Multi-meter HP34420A | 3+23 |
| Function generator HP33120A | 3+10 |
| Oscilloscope HP 54602B | 3+07 |
| PSCx8 Power supply controller | 3+13 |
| Windows PC | |
| Agilent VEE 6.0 | |
| Lab VIEW 6.0 | |

Table-1

*3.1 Connection Test*

When using a Windows PC, the GAC and the measuring devices were connected in the GPIB cables in series. The test program was written in VEE. The following GPIB functions are included for tests: Listener, Talker, Serial poll, SDC, DC, Remote/Local, GET, IFC, and binary data transfer with EOI. The Bus Monitor of the VEE monitors the message transfer on the GPIB. It has been confirmed that the data are smoothly transferred through the GAC. Also, concerning the operation of the application program Agilent VEE: the Panel Drivers of HP34401A, HP34420A, HP33120A, and HP54602B have been confirmed to mount on the Windows PC display correctly and to work without any additional program or change.

*3.2 Transfer Speed*

The transfer speeds were measured under the following two conditions: controller is connected directly to the device (PSCx8); the other is connected through the GAC.
**Test-1:** Loop of 1000 times of (UNL, MTA, LAD, 16 byte DATA with EOI).
**Test-2:** Loop of 1000 times of (UNL, MTA, LAD, 4 byte DATA with EOI, + UNL, MLA, TAD, 21 byte DATA with EOI).
**Test-3:** Serial poll 1000 times.
The test results are shown in Table-2.

| Time (sec) | Test-1 | Test-2 | Test-3 |
|---|---|---|---|
| Direct | 7.299 | 12.22 | 7.236 |
| with GAC | 7.459 | 12.74 | 7.575 |
| +dt (%) | 2.2 | 4.3 | 4.7 |
| Windows PC 450MHz, Agilent VEE 6.01 | | | |

Table-2

The delay time (+dt) % of the data through the SAC on Test-1, Test-2, and Test-3 are, respectively, 2.2%, 4.3%, and 4.7%. The data-transfer speeds of the other devices are slower than the PSCx8, and thus the delay time is smaller.

## 4. CONCLUSION

A GPIB address converter (GAC) has been developed. The function has confirmed that the GAC adds devices with the extended two-byte address function. By using the GAC, it is possible to configure up to 930 ordinal devices equipped with the primary address function on one GPIB.

To improve the SAC transfer speed, two CPUs, H8/3048-(16-bit) and SH7045 (32-bit), are being tested in the SAC.